\documentclass{JHEP3}
\usepackage{graphics}
\newcommand{\Li}[1]{\mathop{\mathrm{Li}}\nolimits_{#1}}
\newcommand{\F}[5]{{}_{#1}F_{#2}\left(\left.\begin{array}{cc}#3\\#4\end{array}\right|#5\right)}

\title{Three-loop HQET vertex diagrams for $B^0$--$\bar{B}^0$ mixing}

\author{Andrey G.~Grozin and Roman N.~Lee\\
Budker Institute of Nuclear Physics, Novosibirsk 630090, Russia\\
E-mail: \email{A.G.Grozin@inp.nsk.su} and \email{R.N.Lee@inp.nsk.su}}

\abstract{Three-loop vertex diagrams in HQET
needed for sum rules for $B^0$--$\bar{B}^0$ mixing are considered.
They depend on two residual energies.
An algorithm of reduction of these diagrams to master integrals
has been constructed.
All master integrals are calculated exactly in $d$ dimensions;
their $\varepsilon$ expansions are also obtained.}

\keywords{NLO Computations, B-Physics}

\begin{document}

\section{Introduction}
\label{S:Intro}

The mass difference $\Delta m$ in $B^0$--$\bar{B}^0$ is determined
in the Standard Model by the matrix element
${<}\bar{B}^0|Q(\mu)|B^0{>}$
of the four-quark operator
\begin{equation}
Q(\mu) = J_\alpha J^\alpha\,,\quad
J^\alpha = \bar{b}_L \gamma^\alpha d_L
\label{Intro:Q}
\end{equation}
(see, e.g., \cite{BBL:96}).
This matrix element is traditionally written as
\begin{equation}
{<}\bar{B}^0|Q(\mu)|B^0{>} =
2 \left(1 + \frac{1}{N_c}\right)
{<}\bar{B}^0|J_\alpha|0{>} {<}0|J^\alpha|B^0{>}
B(\mu)\,,
\label{Intro:B}
\end{equation}
where $N_c$ is the number of colours.
Here the first part of the right-hand side is
the value of the matrix element according to
the naive factorization prescription
(this part does not depend on $\mu$),
and $B(\mu)$ describes violation of this prescription.
The hadronic parameter $B(\mu)$ can only be obtained
by using some non-perturbative method,
such as lattice simulations (see, e.g., \cite{T:07})
or QCD sum rules~\cite{OP:88,KOPP:03,MPP:07}.

In the QCD sum rules approach,
the correlator ${<}jQj{>}$ is investigated,
where $j$ is a current with ${<}B^0|j|0{>}\neq0$
(axial or pseudoscalar).
Contributions to the theoretical expression for this correlator
can be subdivided into two groups:
\begin{equation}
{<}jQj{>} =
2 \left(1 + \frac{1}{N_c}\right) {<}j J_\alpha{>} {<}J^\alpha j{>}
+ {<}jQj{>}_{\mathrm{nf}}\,.
\label{Intro:Corr}
\end{equation}
The first term includes the leading perturbative contribution
plus all corrections (perturbative, vacuum condensates)
to the two two-point correlators ${<}j J_\alpha{>}$, ${<}J^\alpha j{>}$
separately.
It just gives the square of the sum rule for $f_B^2$.
Only the second, non-factorizable part contributes to
the sum rule for $B(\mu)-1$.
Non-factorizable perturbative contributions
first appear at three loops
(one gluon is exchanged between the two two-point correlators).
In general, their calculation is a very difficult three-loop problem with three
energy scales ($m_b^2$, $p_1^2$, $p_2^2$; we suppose that $q^2=0$)
which cannot be solved at present.
Several terms of the expansion in $p_1^2$, $p_2^2$
have been obtained~\cite{KOPP:03}
(this is a much easier single-scale problem).
There are also non-factorizable terms
due to vacuum condensates.

It is also possible to consider sum rules
in the HQET framework (see, e.g., \cite{N:94,G:04}).
The QCD operators $Q$, $j$ can be expressed via HQET operators;
matching coefficients are calculable series in $\alpha_s(m_b)$.
Correlators of HQET operators don't involve the scale $m_b$.
Therefore, no large logarithms appear in perturbative corrections.
On the other hand, derivation and analysis of HQET sum rules
for $1/m_b$ corrections is difficult (though not impossible).
Calculations in HQET are technically easier.
In particular, three-loop diagrams describing the leading perturbative
contribution to the sum rules for $B-1$ involve only two scales ---
two residual energies.
Here we present the method for calculating such diagrams.
Calculation of this perturbative contribution is very desirable,
because it allows one to control the $\mu$-dependence of $B(\mu)-1$.

\section{Reduction}
\label{S:Red}

\FIGURE{
\begin{picture}(108.5,17)
\put(16,9.75){\makebox(0,0){\includegraphics{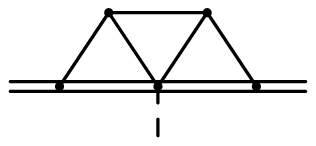}}}
\put(54.25,9.75){\makebox(0,0){\includegraphics{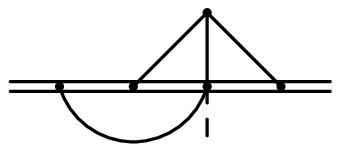}}}
\put(92.5,9.75){\makebox(0,0){\includegraphics{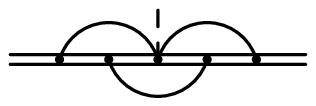}}}
\put(16,0){\makebox(0,0)[b]{$a$}}
\put(54.25,0){\makebox(0,0)[b]{$b$}}
\put(92.5,0){\makebox(0,0)[b]{$c$}}
\end{picture}
\caption{Generic topologies}
\label{F:top}}

\FIGURE{
\begin{picture}(69,12.625)
\put(16,6){\makebox(0,0){\includegraphics{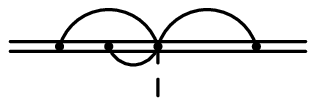}}}
\put(53,6.3125){\makebox(0,0){\includegraphics{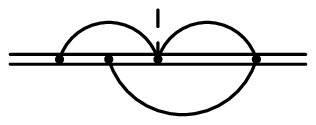}}}
\end{picture}
\caption{Diagrams to which the topology c reduces}
\label{F:pf}}

Non-factorizable three-loop diagrams belong to three topologies
(Fig.~\ref{F:top}).
Four HQET denominators in Fig.~\ref{F:top}$c$ are linearly dependent;
therefore, one heavy line can be killed,
and this diagram reduces to those in Fig.~\ref{F:pf},
which are particular cases of Fig.~\ref{F:top}$b$.

Let the incoming and outgoing residual momenta be $p_{1,2}$.
The scalar integrals depend only on the residual energies
$\omega_{1,2}=p_{1,2}\cdot v$,
where $v$ is the heavy-quark velocity.
In the case $\omega_1=\omega_2$ they reduce
to single-scale HQET integrals~\cite{G:00}
(see also~\cite{G:03,G:08}).

\FIGURE{{}\hspace{20mm}
\begin{picture}(70,33.5)
\put(35,15){\makebox(0,0){\includegraphics{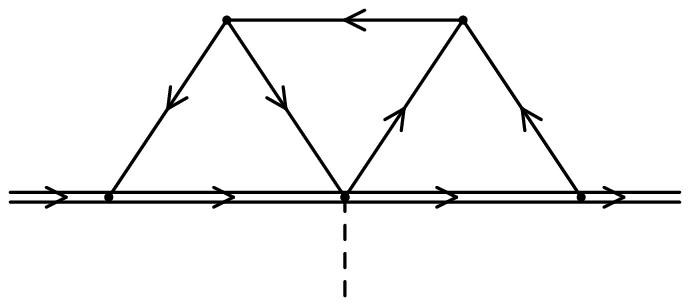}}}
\put(6,10){\makebox(0,0)[t]{$\vphantom{k}\omega_1$}}
\put(64,10){\makebox(0,0)[t]{$\vphantom{k}\omega_2$}}
\put(23,10){\makebox(0,0)[t]{$k_1 v+\omega_1$}}
\put(47,10){\makebox(0,0)[t]{$k_2 v+\omega_2$}}
\put(35,33.5){\makebox(0,0)[t]{$k_3$}}
\put(15,22){\makebox(0,0){$k_1$}}
\put(55,22){\makebox(0,0){$k_2$}}
\put(31,16){\makebox(0,0)[r]{$k_3-k_1$}}
\put(39,16){\makebox(0,0)[l]{$k_3-k_2$}}
\put(23,12){\makebox(0,0)[b]{1}}
\put(47,12){\makebox(0,0)[b]{2}}
\put(35,28){\makebox(0,0)[t]{5}}
\put(19,18){\makebox(0,0)[b]{3}}
\put(51,18){\makebox(0,0)[b]{4}}
\put(32,18){\makebox(0,0)[b]{6}}
\put(38,18){\makebox(0,0)[b]{7}}
\end{picture}\hspace{20mm}{}
\caption{Topology 1}
\label{F:t1}}

We need to consider two topologies.
The first one is (Fig.~\ref{F:t1})
\begin{eqnarray}
&&I_a(n_i;m_j;\omega_1,\omega_2) =
\frac{1}{(i\pi^{d/2})^3} \int
\frac{\prod_j N_j^{m_j}\,d^d k_1\,d^d k_2\,d^d k_3}{\prod_i D_i^{n_i}}\,,
\label{Red:Ia}\\
&&D_1 = -2(k_1\cdot v+\omega_1)\,,\quad
D_2 = -2(k_2\cdot v+\omega_2)\,,\quad
D_3 = -k_1^2\,,\quad
D_4 = -k_2^2\,,
\nonumber\\
&&D_5 = -k_3^2\,,\quad
D_6 = -(k_3-k_1)^2\,,\quad
D_7 = -(k_3-k_2)^2\,,
\nonumber\\
&&N_1 = -2 k_3\cdot v\,,\quad
N_2 = -(k_1-k_2)^2\,,
\nonumber
\end{eqnarray}
where $-i0$ is assumed in all denominators,
$n_i$ and $m_j$ are integer, and $m_j\ge0$.
They can be reduced to master integrals
using integration by parts~\cite{CT:81}.
A \textsf{Mathematica} program
(R.N.~Lee, unpublished, based on~\cite{L:08})
has succeeded in constructing an algorithm
to reduce these scalar integrals to the following
simple master integrals:
\begin{eqnarray}
&&\raisebox{-3.5mm}{\includegraphics{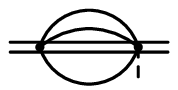}}\,,\qquad
\raisebox{-3.5mm}{\includegraphics{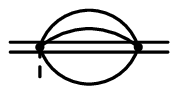}}\,,\qquad
\raisebox{-3.5mm}{\includegraphics{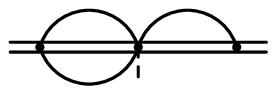}}\,,\qquad
\raisebox{-3.5mm}{\includegraphics{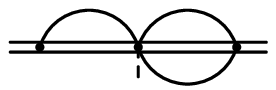}}\,,
\nonumber\\
&&M_1(\omega_1,\omega_2) = \raisebox{-3.5mm}{\includegraphics{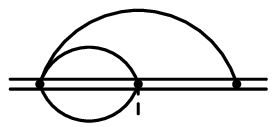}}\,,\qquad
M_1(\omega_2,\omega_1) = \raisebox{-3.5mm}{\includegraphics{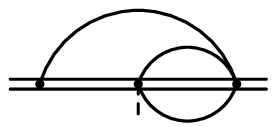}}\,,
\nonumber\\
&&M_2 = \raisebox{-3.5mm}{\includegraphics{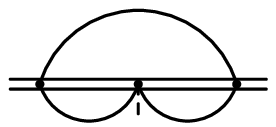}}\,,\qquad
M_2' = \raisebox{-3.5mm}{\includegraphics{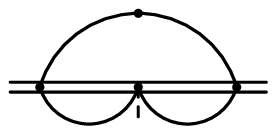}}\,,
\label{Red:Simple}
\end{eqnarray}
and one difficult integral:
\begin{equation}
M_3 = \raisebox{-3.5mm}{\includegraphics{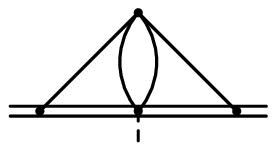}}\,.
\label{Red:I}
\end{equation}

\FIGURE{{}\hspace{20mm}
\begin{picture}(82,35)
\put(41,19){\makebox(0,0){\includegraphics{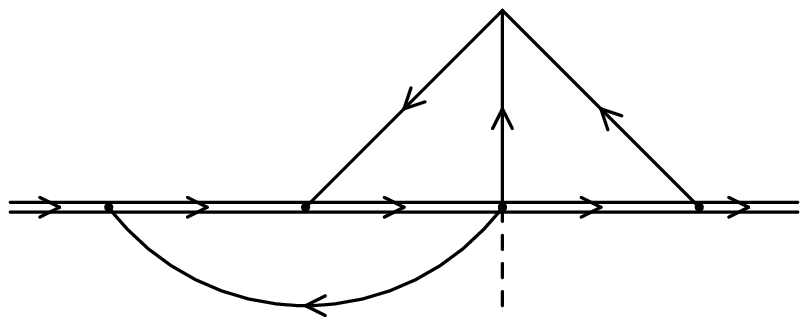}}}
\put(6,15){\makebox(0,0)[b]{$\omega_1$}}
\put(21,15){\makebox(0,0)[b]{$k_1v+\omega_1$}}
\put(76,13){\makebox(0,0)[t]{$\vphantom{k}\omega_2$}}
\put(61,13){\makebox(0,0)[t]{$k_2v+\omega_2$}}
\put(36,13){\makebox(0,0)[t]{$(k_1+k_3)v+\omega_1$}}
\put(31,0){\makebox(0,0)[b]{$k_1$}}
\put(63,26){\makebox(0,0){$k_2$}}
\put(39,26){\makebox(0,0){$k_3$}}
\put(51.5,20){\makebox(0,0)[l]{$k_3-k_2$}}
\put(20,13){\makebox(0,0)[t]{\vphantom{k}1}}
\put(61,15){\makebox(0,0)[b]{2}}
\put(41,15){\makebox(0,0)[b]{3}}
\put(31,5){\makebox(0,0)[b]{4}}
\put(59,23){\makebox(0,0){5}}
\put(43,23){\makebox(0,0){6}}
\put(50,20){\makebox(0,0)[r]{7}}
\end{picture}\hspace{20mm}{}
\caption{Topology 2}
\label{F:t2}}

The second topology is (Fig.~\ref{F:t2})
\begin{eqnarray}
&&I_b(n_i;m_j;\omega_1,\omega_2) =
\frac{1}{(i\pi^{d/2})^3} \int
\frac{\prod_j N_j^{m_j}\,d^d k_1\,d^d k_2\,d^d k_3}{\prod_i D_i^{n_i}}\,,
\label{Red:Ib}\\
&&D_1 = -2(k_1\cdot v+\omega_1)\,,\quad
D_2 = -2(k_2\cdot v+\omega_2)\,,\quad
D_3 = -2((k_1+k_3)\cdot v+\omega_1)\,,
\nonumber\\
&&D_4 = -k_1^2\,,\quad
D_5 = -k_2^2\,,\quad
D_6 = -k_3^2\,,\quad
D_7 = -(k_3-k_2)^2\,,
\nonumber\\
&&N_1 = -(k_1-k_3)^2\,,\quad
N_2 = -(k_1-k_2)^2\,.
\nonumber
\end{eqnarray}
The same program has succeeded in constructing an algorithm
to reduce these scalar integrals
to the same simple master integrals~(\ref{Red:Simple})
and one difficult integral
\begin{equation}
M_4 = \raisebox{-7.9mm}{\includegraphics{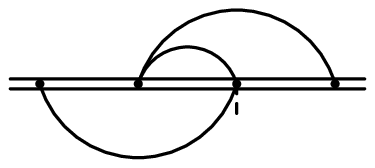}}\,.
\label{Red:J}
\end{equation}

\section{Simple master integrals}
\label{S:Simple}

We consider the below-threshold region $\omega_{1,2}<0$;
expressions for other regions can be obtained by analytical continuation.
The simplest master integrals are single-scale,
or products of single-scale integrals:
\begin{eqnarray}
&&\raisebox{-3.5mm}{\includegraphics{i1.eps}} =
I_3 (-2\omega_1)^{3d-7}\,,
\label{Simple:I1}\\
&&\raisebox{-3.5mm}{\includegraphics{i2.eps}} =
I_1 I_2 (-2\omega_1)^{2d-5} (-2\omega_2)^{d-3}\,,
\end{eqnarray}
where the $n$-loop HQET sunset is
\begin{equation}
I_n = \Gamma(2n+1-nd) \Gamma^n\left({\textstyle\frac{d}{2}}-1\right)\,.
\label{Simple:In}
\end{equation}

Several master integrals reduce to the one-loop vertex
with two residual energies
\begin{eqnarray}
&&\raisebox{-5.25mm}{\begin{picture}(32,16.5)
\put(16,7.25){\makebox(0,0){\includegraphics{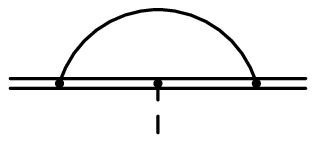}}}
\put(3.5,4.5){\makebox(0,0)[t]{$\omega_1$}}
\put(28.5,4.5){\makebox(0,0)[t]{$\omega_2$}}
\put(11,4.5){\makebox(0,0)[t]{$n_1$}}
\put(21,4.5){\makebox(0,0)[t]{$n_2$}}
\put(16,16.5){\makebox(0,0)[t]{$n_3$}}
\end{picture}} =
I(n_1,n_2,n_3;\omega_1,\omega_2) =
\frac{1}{i\pi^{d/2}} \int
\frac{d^d k}{D_1^{n_1} D_2^{n_2} D_3^{n_3}}\,,
\label{Simple:Vdef}\\
&&D_1 = -2(k\cdot v+\omega_1)\,,\quad
D_2 = -2(k\cdot v+\omega_2)\,,\quad
D_3 = -k^2\,.
\nonumber
\end{eqnarray}
It is~\cite{BBG:93}
\begin{eqnarray}
I(n_1,n_2,n_3;\omega_1,\omega_2) &=& I(n_1+n_2,n_3)\,
\F{2}{1}{n_1,n_1+n_2+2n_3-d}{n_1+n_2}{1-\frac{1}{x}}
\nonumber\\
&&{}\times(-2\omega_2)^{d-n_1-n_2-2n_3}\,,
\label{Simple:V}
\end{eqnarray}
where the HQET two-point integral is
\begin{equation}
I(n_1,n_2) =
\frac{\Gamma(n_1+2n_2-d) \Gamma\left(\frac{d}{2}-n_2\right)}%
{\Gamma(n_1) \Gamma(n_2)}\,,
\label{Simple:I}
\end{equation}
and
\begin{equation}
x = \frac{\omega_2}{\omega_1}\,.
\label{Simple:x}
\end{equation}
Naturally,
\begin{eqnarray*}
&&I(n_1,n_2,n_3;\omega_1,\omega_2) = I(n_2,n_1,n_3;\omega_2,\omega_1)\,,\\
&&I(n_1,n_2,n_3;\omega,\omega) = I(n_1+n_2,n_3) (-2\omega)^{d-n_1-n_2-2n_3}\,.
\end{eqnarray*}
Later we shall also need
\begin{eqnarray}
&&I(n_1,n_2,n_3;\omega,0) = I_0(n_1,n_2,n_3) (-2\omega)^{d-n_1-n_2-2n_3}\,,
\nonumber\\
&&I_0(n_1,n_2,n_3) =
\frac{\Gamma\left(\frac{d}{2}-n_3\right) \Gamma(d-n_2-2n_3)
\Gamma(n_1+n_2+2n_3-d)}{\Gamma(n_1) \Gamma(n_3) \Gamma(d-2n_3)}\,.
\label{Simple:I0}
\end{eqnarray}
Several ways to derive~(\ref{Simple:V}) are discussed in~\cite{G:08}.

Using this integral, we easily obtain
\begin{eqnarray}
&&M_1(\omega_1,\omega_2)
= I_2 I(5-2d,1,1;\omega_1,\omega_2)
= I_2 I(1,5-2d,1;\omega_2,\omega_1)\,,
\label{Simple:M1}\\
&&M_2(\omega_1,\omega_2)
= I_1^2 I(3-d,3-d,1;\omega_1,\omega_2)\,,
\label{Simple:M2}\\
&&M_2'(\omega_1,\omega_2)
= I_1^2 I(3-d,3-d,2;\omega_1,\omega_2)\,.
\label{Simple:I4a}
\end{eqnarray}

\section{Master integral $M_4$}
\label{S:J}

We were able to calculate a more general integral
\begin{equation}
J(n_1,n_2,n_3,n_4,n_5;\omega_1,\omega_2) =
\raisebox{-9.7mm}{\begin{picture}(42,21)
\put(21,10.5){\makebox(0,0){\includegraphics{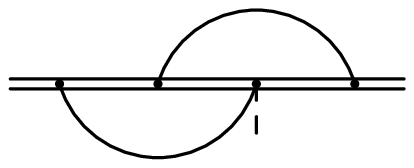}}}
\put(3.5,11.5){\makebox(0,0)[b]{$\omega_1$}}
\put(38.5,9.5){\makebox(0,0)[t]{$\omega_2$}}
\put(11,11.5){\makebox(0,0)[b]{$n_1$}}
\put(21,11.5){\makebox(0,0)[b]{$n_3$}}
\put(31,9.5){\makebox(0,0)[t]{$n_2$}}
\put(16,0){\makebox(0,0)[b]{$n_4$}}
\put(26,21){\makebox(0,0)[t]{$n_5$}}
\end{picture}}\,.
\label{J:def}
\end{equation}
Substituting~(\ref{Simple:V}) for the left one-loop vertex subdiagram,
we have
\begin{eqnarray*}
&&\frac{I(n_1+n_3,n_4)}{i\pi^{d/2}} (-2\omega_1)^{d-n_1-n_3-2n_4}\\
&&{}\times\int \frac{d k_0\,d^{d-1}\vec{k}}%
{(-k^2)^{n_5} (-2(k_0+\omega_2))^{n_2}}\,
\F{2}{1}{n_1,n_1+n_3+2n_4-d}{n_1+n_3}{-\frac{k_0}{\omega_1}}\,.
\end{eqnarray*}
Then we perform Wick rotation $k_0=ik_{E0}$
and take the $d^{d-1}\vec{k}$ integral.
The integrand has a cut from 0 to $+i\infty$;
we deform the integration contour around this cut
($k_{E0}=i(-\omega_2)z$):
\begin{eqnarray*}
&&\frac{I(n_1+n_3,n_4) \Gamma\left(n_5-\frac{d-1}{2}\right)}%
{\pi^{1/2} 2^{d-2n_5-1} \Gamma(n_5)}
\cos\left[\pi\left({\textstyle\frac{d}{2}}-n_5\right)\right]
(-2\omega_1)^{d-n_1-n_3-2n_4} (-2\omega_2)^{d-n_2-2n_5}\\
&&{}\times\int_0^\infty \frac{d z\,z^{d-2n_5-1}}{(z+1)^{n_2}}\,
\F{2}{1}{n_1,n_1+n_3+2n_4-d}{n_1+n_3}{-xz}\,.
\end{eqnarray*}
This integral can be calculated in terms of two ${}_3F_2$ functions,
and we arrive at
\begin{eqnarray}
&&J(n_1,n_2,n_3,n_4,n_5;\omega_1,\omega_2) =
\frac{\Gamma\left(\frac{d}{2}-n_4\right) \Gamma\left(\frac{d}{2}-n_5\right)}%
{\Gamma(n_4) \Gamma(n_5)}
\nonumber\\
&&{}\times\Biggl[
\frac{\Gamma(n_1+n_3+2n_4-d) \Gamma(n_2+2n_5-d)}%
{\Gamma(n_2) \Gamma(n_1+n_3)}
\nonumber\\
&&\qquad{}\times
\F{3}{2}{n_3,n_1+n_3+2n_4-d,d-2n_5}{n_1+n_3,d-n_2-2n_5+1}{x}
x^{d-n_2-2n_5}
\nonumber\\
&&\qquad{} +
\frac{\Gamma(d-n_2-2n_5) \Gamma(n_2+n_3+2n_5-d)
\Gamma(n_1+n_2+n_3+2n_4+2n_5-2d)}%
{\Gamma(n_3) \Gamma(d-2n_5) \Gamma(n_1+n_2+n_3+2n_5-d)}
\nonumber\\
&&\qquad\qquad{}\times
\F{3}{2}{n_2,n_2+n_3+2n_5-d,n_1+n_2+n_3+2n_4+2n_5-2d}%
{n_2+2n_5-d+1,n_1+n_2+n_3+2n_5-d}{x}
\Biggr]
\nonumber\\
&&{}\times(-2\omega_1)^{2d-n_1-n_2-n_3-2n_4-2n_5}\,.
\label{J:J}
\end{eqnarray}
Trivial cases are reproduced:
\begin{eqnarray*}
&&J(n_1,n_2,0,n_4,n_5;\omega_1,\omega_2) =
I(n_1,n_4) I(n_2,n_5)
(-2\omega_1)^{d-n_1-2n_4} (-2\omega_2)^{d-n_2-2n_5}\,,\\
&&J(n_1,0,n_3,n_4,n_5;\omega_1,\omega_2) =
I(n_3,n_5) I(n_1+n_3+2n_5-d,n_4)
(-2\omega_1)^{2d-n_1-n_3-2n_4-2n_5}\,.
\end{eqnarray*}
At $\omega_1=\omega_2$, the single-scale integral~\cite{G:00,G:03}
is reproduced (its derivation is also discussed in~\cite{G:08}).

Now it is easy to write down the master integral~(\ref{Red:J})
\begin{equation}
M_4(\omega_1,\omega_2) = I_1 J(1,1,3-d,1,1;\omega_1,\omega_2)\,.
\label{J:master}
\end{equation}
Note that the first ${}_3F_2$ function in Eq.~(\ref{J:J})
turns into ${}_2F_1$ when one substitutes $n_2=1$ in order to obtain $M_4$.

\section{Master integral $M_3$}
\label{S:I}

This integral can be expressed as
\begin{equation}
M_3(\omega_1,\omega_2) = G_1
I(1,1,1,1,2-{\textstyle\frac{d}{2}};\omega_1,\omega_2)
\label{I:M3}
\end{equation}
via the two-loop integral
\begin{eqnarray}
&&I(n_1,n_2,n_3,n_4,n_5;\omega_1,\omega_2) =
\raisebox{-5.2mm}{\begin{picture}(42,22)
\put(21,11){\makebox(0,0){\includegraphics{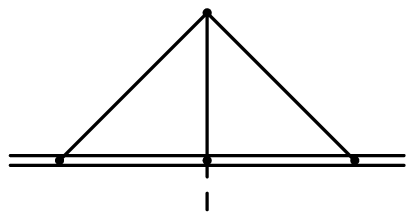}}}
\put(3.5,5){\makebox(0,0)[t]{$\omega_1$}}
\put(38.5,5){\makebox(0,0)[t]{$\omega_2$}}
\put(13.5,5){\makebox(0,0)[t]{$n_1$}}
\put(28.5,5){\makebox(0,0)[t]{$n_2$}}
\put(12,15){\makebox(0,0){$n_3$}}
\put(30,15){\makebox(0,0){$n_4$}}
\put(22,11){\makebox(0,0)[l]{$n_5$}}
\end{picture}} =
\frac{1}{(i\pi^{d/2})^2} \int \frac{d^d k_1\,d^d k_2}{\prod_i D_i^{n_i}}\,,
\label{I:I}\\
&&D_1 = - 2 (k_1\cdot v + \omega_1)\,,\quad
D_2 = - 2 (k_2\cdot v + \omega_2)\,,
\nonumber\\
&&D_3 = - k_1^2\,,\quad
D_4 = - k_2^2\,,\quad
D_5 = - (k_1-k_2)^2
\nonumber
\end{eqnarray}
with non-integer $n_5$, where
\begin{equation}
G_n = \frac{\Gamma\left(n+1-n\frac{d}{2}\right)
\Gamma^{n+1}\left(\frac{d}{2}-1\right)}%
{\Gamma\left((n+1)\left(\frac{d}{2}-1\right)\right)}
\label{I:G}
\end{equation}
is the $n$-loop massless sunset.

In order to express $M_3$ in closed form,
we can use the method of differential equations~\cite{Ko:91,Re:97}.
The differential equation for this master integral can be obtained
by differentiating it with respect to $\omega_1$
and then applying the reduction rules
obtained by the \textsf{Mathematica} program.
It reads
\begin{eqnarray}
&&\omega_1 \frac{\partial M_3(\omega_1,\omega_2)}{\partial\omega_1}
= \frac{3d-10}{2} M_3(\omega_1,\omega_2) + H(\omega_1,\omega_2)\,,
\label{I:DE}\\
&&H(\omega_1,\omega_2)
= \frac{2d-5}{2(\omega_1-\omega_2)^2} M_1(\omega_2,\omega_1)
- \frac{(3d-7) [(3d-8)\omega_1-(5d-14)\omega_2]}%
{8 (d-3) \omega_2^2 (\omega_1-\omega_2)^2}
I_3 (-2\omega_2)^{3d-7}
\nonumber\\
&&\hphantom{H(\omega_1,\omega_2)}{}
- \frac{2d-5}{2(\omega_1-\omega_2)^2} M_1(\omega_1,\omega_2)
+ \frac{(3d-7) [(3d-8)\omega_2-(5d-14)\omega_1]}%
{8 (d-3) \omega_1^2 (\omega_1-\omega_2)^2}
I_3 (-2\omega_1)^{3d-7}\,.
\nonumber
\end{eqnarray}
Using the explicit expressions for the simple master integrals,
it is easy to check that singularities at $\omega_1=\omega_2$
cancel in $H$ separately on the second and third lines in Eq. (\ref{I:DE}).

The general solution of this differential equation has the form
\[
M_3(\omega_1,\omega_2) = M_0(\omega_1)
\left[C + \int_{-\infty}^{\omega_1} d \omega\, M_0^{-1}(\omega) H(\omega,\omega_2)
\right]\,,
\]
where
\[
M_0(\omega)=(-2\omega)^{3d/2-5}
\]
is the solution of the homogeneous part of the equation~(\ref{I:DE}).
In order to fix the constant $C$,
we consider the asymptotics of $M_3(\omega_1,\omega_2)$
when $\omega_1\to-\infty$~\cite{BL:06}.
Using the method of expansion by regions (see~\cite{S:02}),
it is easy to determine that there is no $\mathcal{O}(\omega_1^{3d/2-5})$ term
in the asymptotics.
Thus, $C=0$, and we obtain
\begin{eqnarray}
&& M_3(\omega_1,\omega_2) =
2 (-2\omega_1)^{3d/2-5} (-2\omega_2)^{3d/2-5}
\Gamma^3\left({\textstyle\frac{d}{2}}-1\right) \Gamma(8-3d)
\int_{1/x}^\infty \frac{dy}{(y-1)^{2}}
\nonumber\\
&&{}\times
\Biggl\{y^{4-3d/2} \left[ \F{2}{1}{1,8-3d}{6-2d}{1-y}
- 1 - \frac{8-3d}{6-2d} (1-y) \right]
\nonumber\\
&&\qquad{} - y^{3d/2-4} \left[ \F{2}{1}{1,8-3d}{6-2d}{1-\frac{1}{y}}
- 1 - \frac{8-3d}{6-2d} \left(1 - \frac{1}{y} \right) \right]
\Biggr\}\,.
\label{eq:int}
\end{eqnarray}
Note that the rational terms in brackets
are the two first terms of expansion of the corresponding ${}_2F_1$
with respect to its argument.
Now, using the parametrization
\begin{eqnarray}
&&\F{2}{1}{1,8-3d}{6-2d}{1-t}
- 1 - \frac{8-3d}{6-2d} (1-t)
\nonumber \\
&&{} = \frac{\Gamma(6-2d)}{\Gamma(8-3d) \Gamma(d-2)}
\int_0^\infty ds\,s^{7-3d} (1+s)^{2d-5}
\left(\frac{1}{1+st} - \frac{1}{1+s} - \frac{s(1-t)}{(1+s)^{2}}\right)\,,
\label{eq:param}
\end{eqnarray}
we can take the integrals first over $y$ and then over $s$.
Finally, we obtain
\begin{eqnarray}
M_3(\omega_1,\omega_2) &=&
4 (-2\omega_1)^{3d-10} \Gamma^3\left({\textstyle\frac{d}{2}}-1\right)
\nonumber\\
&&{}\times\Biggl[\frac{\Gamma(8-3d)}{2(d-3)} x^{3d-9}\,
\F{3}{2}{1,d-2,\frac{3}{2}d-4}{\frac{3}{2}d-3,3d-8}{x}
\nonumber\\
&&\qquad{} + \frac{3\Gamma(9-3d)}{2(d-3)(3d-10)}\,
\F{3}{2}{1,10-3d,5-\frac{3}{2}d}{6-\frac{3}{2}d,4-d}{x}
\nonumber\\
&&\qquad{} + \frac{\pi\Gamma(6-2d)}{(3d-10)\Gamma(d-2)\sin(3\pi d)}\,
\F{2}{1}{5-\frac{3}{2}d,7-2d}{6-\frac{3}{2}d}{x}
\nonumber\\
&&\qquad{} + \frac{\pi\Gamma(6-2d)}{(d-4)\Gamma(d-2)\sin(\pi d)} x^{d-3}\,
\F{2}{1}{2-\frac{d}{2},7-2d}{3-\frac{d}{2}}{x}
\Biggr]\,.
\label{eq:M1}
\end{eqnarray}
It follows from the analyticity of $M_3(\omega_1,\omega_2)$
in the region $\omega_{1,2}<0$ that the above expression
is analytical in the interval $x\in(0,+\infty)$.
In particular, branching singularities at $x=1$ cancel.

The integral $M_3$ is a symmetric function of its arguments.
This symmetry can be made explicit if we rewrite the integral over $y$
of the terms in the last line of~(\ref{eq:int}) as follows:
\[
\int_{1/x}^\infty dy = \int_0^\infty dy
- \int_x^\infty d(1/y)y^2\,,
\]
and make the replacement $y\to 1/y$ in the second integral.
Then, using the same parametrization~(\ref{eq:param}), we obtain
\begin{eqnarray}
M_3(\omega_1,\omega_2) &=&
(-2\omega_1)^{3d/2-5} (-2\omega_2)^{3d/2-5} \Gamma^3(d/2-1)
\nonumber\\
&&{}\times \Biggl[
\frac{\Gamma\left(\frac{3}{2}d-4\right) \Gamma^2\left(5-\frac{3}{2}d\right)
\Gamma\left(2-\frac{d}{2}\right)}{(d-3) \Gamma(d-2)}
\nonumber\\
&&\qquad{} + 2 \frac{\Gamma(8-3d)}{d-3} x^{4-3d/2}\,
\F{3}{2}{1,d-2,\frac{3}{2}d-4}{\frac{3}{2}d-3,3d-8}{\frac{1}{x}}
\nonumber\\
&&\qquad{} + \frac{4 \pi \Gamma(6-2d) x^{3d/2-5}}%
{(3d-10) \Gamma(d-2) \sin(3\pi d)}\,
\F{2}{1}{5-\frac{3}{2}d,7-2d}{6-\frac{3}{2}d}{\frac{1}{x}}
\nonumber\\
&&\qquad{} + 2\frac{\Gamma(8-3d)}{d-3} x^{3 d /2-4}\,
\F{3}{2}{1,d-2,\frac{3}{2}d-4}{\frac{3}{2}d-3,3d-8}{x}
\nonumber\\
&&\qquad{} + \frac{4 \pi \Gamma(6-2d) x^{5-3d/2}}%
{(3d-10) \Gamma(d-2) \sin(3\pi d)}\,
\F{2}{1}{5-\frac{3}{2}d,7-2d}{6-\frac{3}{2}d}{x}
\Biggr]\,.
\label{eq:M2}
\end{eqnarray}

We have performed two crucial checks of the above expressions for $M_3$.
The first check is due to the fact that at  $\omega_1=\omega_2$ the integral $M_3$ reduces
to the known single-scale integral~\cite{BB:94}. Though our representations
do not literally coincide with those in \cite{BB:94},
we have been able to check the perfect numerical agreement.

\FIGURE{
\begin{picture}(136,25)
\put(21,14){\makebox(0,0){\includegraphics{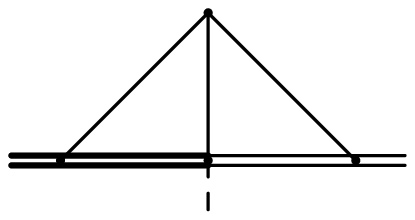}}}
\put(68,14){\makebox(0,0){\includegraphics{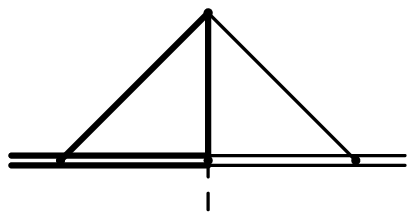}}}
\put(115,14){\makebox(0,0){\includegraphics{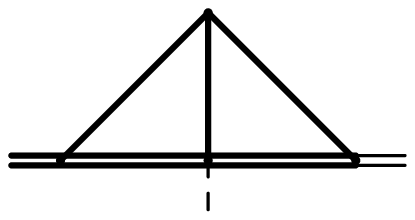}}}
\put(22,14){\makebox(0,0)[l]{$n$}}
\put(69,14){\makebox(0,0)[l]{$n$}}
\put(116,14){\makebox(0,0)[l]{$n$}}
\put(21,0){\makebox(0,0)[b]{$a$}}
\put(68,0){\makebox(0,0)[b]{$b$}}
\put(115,0){\makebox(0,0)[b]{$c$}}
\end{picture}
\caption{Regions: thick lines are hard (momenta $\sim\omega_1$),
thin lines are soft (momenta $\sim\omega_2$)}
\label{F:Regions}}

The asymptotics of $M_3$~(\ref{eq:M1}) at $x\to0$
can be also obtained by using the method of regions~\cite{S:02}
for $I(1,1,1,1,n;\omega_1,\omega_2)$ with $n=2-d/2$~(\ref{I:M3}).
There are 3 regions shown in Fig.~\ref{F:Regions}.
The region $a$ gives the first term in~(\ref{eq:M1});
$b$ --- the fourth term;
and $c$ --- the second and the third ones
(this is clear from the powers of $\omega_{1,2}$).

In the region $a$, we expand $1/D_1$~(\ref{I:I}) in $k_1\cdot v$.
Then we calculate the left massless loop (lines 3 and 5)
with the numerator $(k_1\cdot v)^l$
(see, e.g., \cite{S:02}, eqs.~(A.11), (A.12)).
In the numerator of the remaining HQET integral,
powers of $2k_2\cdot v$ may be replaced by powers of $-2\omega_2$,
because integrals in which the denominator $D_2$ cancels are zero.
We obtain a series in $x$ whose coefficients are finite sums.
We have checked that a few terms in this series agree with
the expansion of the first term in~(\ref{eq:M1}).

In the region $b$, we expand $1/D_5^n$~(\ref{I:I}) in $k_2$.
Then we calculate the left (hard) HQET loop with a numerator
(see~\cite{G:00}, eq.~(2.13)),
and finally the right (soft) HQET loop
(it also has numerators).
The coefficients of the resulting series are finite sums.
We have checked that a few terms in this series
agree with the expansion of the fourth term in~(\ref{eq:M1}).

In the region $c$, we expand $1/D_2$~(\ref{I:I}) in $\omega_2$:
\[
(-2\omega_1)^{2d-2n-6} \sum_{l=0}^\infty I_0(1,l+1,1,1,n) (-x)^l\,,
\]
where
\[
I(n_1,n_2,n_3,n_4,n_5;\omega,0) = I_0(n_1,n_2,n_3,n_4,n_5)
(-2\omega)^{2d-n_1-n_2-2n_3-2n_4-2n_5}\,.
\]
Using integration by parts, we obtain
\[
I_0(1,l+1,1,1,n) = n
\frac{I_0(1,l+1,1,0,n+1)-I_0(1,l+1,0,1,n+1)}{d-n-l-3}\,,
\]
where
\begin{eqnarray*}
&&I_0(n_1,n_2,0,n_4,n_5) = I(n_1,n_5) I_0(n_1+2n_5-d,n_2,n_4)\,,\\
&&I_0(n_1,n_2,n_3,0,n_5) = I(n_2,n_5) I_0(n_1,n_2+2n_5-d,n_3)
\end{eqnarray*}
(see~(\ref{Simple:I0})).
The contribution of the region $c$ is thus
\begin{eqnarray*}
&&\frac{\Gamma\left(\frac{d}{2}-1\right)
\Gamma\left(\frac{d}{2}-n-1\right) \Gamma(2n+6-2d)}%
{(d-n-3) \Gamma(n)}\\
&&{}\times\Biggl[
\frac{\Gamma(2d-2n-5) \Gamma(2n+3-d)}{\Gamma(d-2)}\,
\F{2}{1}{n+3-d,2n+3-d}{n+4-d}{x}\\
&&\hphantom{{}\times\Biggl[\Biggr.}{}
- \frac{1}{d-3}\,
\F{3}{2}{1,n+3-d,2n+6-2d}{4-d,n+4-d}{x}
\Biggr]\,.
\end{eqnarray*}
Substituting $n=2-d/2$ and multiplying by $G_1$ (see~(\ref{I:M3})),
we reproduce the second and the third terms in~(\ref{eq:M1}).

\section{Conclusion}
\label{S:Conc}

We have considered scalar loop integrals needed for
the perturbative part of HQET sum rules for $B-1$.
The sum rules will be considered in a future publication.
The width difference $\Delta\Gamma$ involves matrix elements
of four-quark operators similar to~(\ref{Intro:Q})
but with different Dirac structures.
In higher orders in $1/m_b$, similar operators
involving derivatives appear.
Matrix elements of such operators can also be estimated
using HQET sum rules
(operators with derivatives are very difficult
for lattice simulations).

More general classes of three-loop HQET vertex diagrams can be analyzed
using the same method.
Master integrals calculated here will be useful for such an analysis.

We are grateful to A.A.~Pivovarov for discussions of HQET sum rules
for $B^0$--$\bar{B}^0$ mixing.

\appendix
\section{Expansions in $\varepsilon$}
\label{S:App}

We use the \textsf{Mathematica} package \textsf{HypExp}~\cite{HM:06}
to expand the master integrals in $\varepsilon$ ($d=4-2\varepsilon$):
\begin{eqnarray}
M_1 &=& \frac{\Gamma^3(1-\varepsilon) \Gamma(1+6\varepsilon)}%
{72 \varepsilon^2 (1-2\varepsilon) (1-3\varepsilon) (2-3\varepsilon)
(3-4\varepsilon) (1-6\varepsilon)}
\biggl\{ 3 x (1-x)^3
\nonumber\\
&&{} - \frac{1}{2} \Bigl[
36 x (1-x)^3 \log x
- 6 + 71 x - 141 x^2 + 105 x^3 - 27 x^4
\Bigr] \varepsilon
\nonumber\\
&&{} - \frac{1}{2} (1-x) \Bigl[
18 x (1-x)^2 \left( 8 L(x) - 4 \log^2 x - 9 \log x \right)
- 4 + 63 x - 78 x^2 + 21 x^3
\Bigr] \varepsilon^2
\nonumber\\
&&{} + (1-x) \Bigl[
9 x (1-x)^2 \bigl( 48 \Li3(1-x) + 16 \Li3(1-x^{-1}) - 4 \log^3 x + 36 L(x)
\nonumber\\
&&\qquad{}
- 18 \log^2 x + 7 \log x \bigr)
+ 2 (2 - 54 x + 69 x^2 - 18 x^3)
\Bigr] \varepsilon^3
+ \cdots \biggr\} (-2\omega_1)^{4-6\varepsilon}\,,
\label{App:M1}\\
M_2 &=& \frac{(1-4\varepsilon) \Gamma^3(1-\varepsilon)
\Gamma^2(1+2\varepsilon) \Gamma(1+6\varepsilon)}%
{36 \varepsilon^2 (1-2\varepsilon)^2 (1-3\varepsilon) (2-3\varepsilon)
(1-6\varepsilon) \Gamma(1+4\varepsilon)}
\biggl\{
x^2 - \frac{1}{2} (1-x)^2 (1+x^2)
\nonumber\\
&&{} - \frac{3}{2} (1-x)^2 \Bigl[ (1-x) (1+x) \log x + x \Bigr] \varepsilon
\nonumber\\
&&{} - \frac{3}{4} \Bigl[
8 (1-x)^3 (1+x) L(x)
+ (1-2x-2x^3+x^4) \log^2 x
\nonumber\\
&&\hphantom{{}-\frac{3}{4}\Bigl[\biggr.}{}
- 2 x (1-x) (1+x) \log x
+ 8 x (1-x)^2 \Bigr] \varepsilon^2
\nonumber\\
&&{} + \frac{1}{4} \Bigl[
96 x^3 (2-x) \Li3(1-x) - 96 (1-2x) \Li3(1-x^{-1})
\nonumber\\
&&\hphantom{{}+\frac{1}{4}\Bigl[\biggr.}{}
+ 24 (1-2x-2x^3+x^4) L(x) \log x
+ (1-x)^3 (1+x) \log^3 x
\nonumber\\
&&\hphantom{{}+\frac{1}{4}\Bigl[\biggr.}{}
+ 24 x (1-x) (1+x) (L(x) + \log x)
\nonumber\\
&&\hphantom{{}+\frac{1}{4}\Bigl[\biggr.}{}
- 3 x (1-x)^2 \left( 3 \log^2 x + 32 \right)
\Bigr] \varepsilon^3
+ \cdots \biggr\} (-2\omega_1)^{4-6\varepsilon} x^{-3\varepsilon}\,,
\label{App:M2}\\
M_2' &=& - \frac{(1-4\varepsilon) \Gamma^3(1-\varepsilon)
\Gamma^2(1+2\varepsilon) \Gamma(1+6\varepsilon)}%
{6 \varepsilon^3 (1-2\varepsilon) (1-3\varepsilon)
(1-6\varepsilon) \Gamma(1+4\varepsilon)}
\biggl\{ x
+ \frac{1}{2} (1-x)^2 \varepsilon
\nonumber\\
&&{} + \frac{1}{2} \Bigl[ 3 x \log^2 x + 3 (1-x) (1+x) \log x + 4 (1-x)^2
\Bigr] \varepsilon^2
\nonumber\\
&&{} + \frac{1}{4} \Bigl[
48 x \left( 2 \Li3(1-x) + 2 \Li3(1-x^{-1}) - L(x) \log x \right)
\nonumber\\
&&\hphantom{{}+\frac{1}{4}\Bigl[\biggr.}{}
+ 24 (1-x) (1+x) (L(x) + \log x)
+ (1-x)^2 \left( 3 \log^2 x + 32 \right)
\Bigr] \varepsilon^3
\nonumber\\
&&{}
+ \cdots \biggr\} (-2\omega_1)^{2-6\varepsilon} x^{-3\varepsilon}\,,
\label{App:M2p}\\
M_3 &=& \frac{\Gamma^3(1-\varepsilon) \Gamma(1+6\varepsilon)}%
{36 \varepsilon^3 (1-2\varepsilon)^2 (1-3\varepsilon) (2-3\varepsilon)
(1-6\varepsilon)}
\biggl\{ 6 x
- 3 (1 + 13 x + x^2) \varepsilon
\nonumber\\
&&{} - \frac{1}{2} \Bigl[
2 x \left( 9 \log^2 x + 16 \pi^2 \right)
+ 18 (1-x) (1+x) \log x - 9 (1+x)^2
\Bigr] \varepsilon^2
\nonumber\\
&&{} - \frac{1}{2} \Bigl[
48 x \left( 12 \Li3(1-x) + 12 \Li3(1-x^{-1}) - 6 L(x) \log x
- 28 \zeta(3) - 5 \pi^2 \right)
\nonumber\\
&&\hphantom{{}-\frac{1}{2}\Bigl[\Bigr.}{}
+ 9 (1-x) (1+x) \left( 16 L(x) - 3 \log x \right)
- 9 (1 + 13 x + x^2) \log^2 x
\nonumber\\
&&\hphantom{{}-\frac{1}{2}\Bigl[\Bigr.}{}
- 12 (1 - 15 x + x^2)
\Bigr] \varepsilon^3
+ \cdots \biggr\} (-2\omega_1)^{2-6\varepsilon} x^{-3\varepsilon}\,,
\label{App:M3}\\
M_4 &=& \frac{\Gamma^3(1-\varepsilon) \Gamma(1+6\varepsilon)}%
{24 \varepsilon^3 (1-2\varepsilon)^3 (1-3\varepsilon) (1-6\varepsilon)}
\biggl\{ x^2
- \frac{1}{2} x \Bigl[ 6 x \log x - 1 + 18 x + x^2 \Bigr] \varepsilon
\nonumber\\
&&{} + \frac{1}{6} \Bigl[
6 x^2 \left( 3 \log^2 x - 2 \pi^2 \right)
- 18 x (1 - 9 x - x^2) \log x
+ 2 - 45 x + 96 x^2 + 15 x^3
\Bigr] \varepsilon^2
\nonumber\\
&&{} - \Bigl[ 2 x^2 \left( 24 \Li3(1-x) + 24 \Li3(1-x^{-1}) - 12 L(x) \log x
- 4 \pi^2 \log x - 60 \zeta(3) - 9 \pi^2 \right)
\nonumber\\
&&\hphantom{{}-\Bigl[\Bigr.}{} + 12 x (1-x) (1+x) L(x)
- 3 x (2 - 9 x - 2 x^2) \log^2 x
- 3 x (5 - 18 x - 5 x^2) \log x
\nonumber\\
&&\hphantom{{}-\Bigl[\Bigr.}{} + x (7 + 2 x - x^2)
\Bigr] \varepsilon^3 + \cdots \biggr\} (-2\omega_1)^{3-6\varepsilon}\,,
\label{App:M4}
\end{eqnarray}
where
\[
L(x) = - L(x^{-1}) =
\Li2(1-x) + \frac{1}{4} \log^2 x\,.
\]
As it was mentioned above, all the master integrals
are analytical in $x\in(0,+\infty)$, and hence the coefficients in
the expansions~(\ref{App:M1})--(\ref{App:M4}) are analytical, too.
It is easy to see that $M_2$, $M_2'$, $M_3$ are
symmetric with respect to $\omega_1\leftrightarrow\omega_2$.
The series~(\ref{App:M3}), (\ref{App:M4}) at
$\omega_1=\omega_2$ coincide with the expansions
of the single-scale integrals~\cite{BB:94,G:00,G:03}.

\end{document}